\documentclass[10pt]{article}
\usepackage{graphicx}
\usepackage{amsmath}
\usepackage{amssymb}
\usepackage{caption2}
\begin{document}
\bibliographystyle{prsty}
\begin{center}
{\large {\bf \sc{  Reanalysis of the $X(3915)$, $X(4500)$ and $X(4700)$ with  QCD sum rules }}} \\[2mm]
Zhi-Gang  Wang \footnote{E-mail: zgwang@aliyun.com.  }     \\
 Department of Physics, North China Electric Power University, Baoding 071003, P. R. China
\end{center}

\begin{abstract}
In this article, we study the   $C\gamma_5\otimes \gamma_5C$ type and $C\otimes C$ type scalar $cs\bar{c}\bar{s}$ tetraquark states with the QCD sum rules  by calculating the contributions of the vacuum condensates up to dimension 10 in a consistent way.  The  ground state masses $M_{C\gamma_5\otimes \gamma_5C}=3.89\pm 0.05\,\rm{GeV}$  and $M_{C\otimes C}=5.48\pm0.10\,\rm{GeV}$ support assigning the $X(3915)$ to be the ground state $C\gamma_5\otimes \gamma_5C$ type tetraquark state with $J^{PC}=0^{++}$, but  do not support  assigning the $X(4700)$ to be the ground state $C\otimes C$ type $cs\bar{c}\bar{s}$ tetraquark state with $J^{PC}=0^{++}$.
 Then we tentatively assign the $X(3915)$ and $X(4500)$ to be the  1S and 2S  $C\gamma_5\otimes \gamma_5C$ type  scalar $cs\bar{c}\bar{s}$ tetraquark states respectively,  and obtain the 1S mass $M_{\rm 1S}=3.85^{+0.18}_{-0.17}\,\rm{GeV}$ and 2S mass  $M_{\rm 2S}=4.35^{+0.10}_{-0.11}\,\rm{GeV}$ from the QCD sum rules, which support assigning the $X(3915)$ to be the 1S $C\gamma_5\otimes \gamma_5C$ type tetraquark state, but  do not support  assigning the $X(4500)$ to be the 2S $C\gamma_5\otimes \gamma_5C$ type tetraquark state.
\end{abstract}



\section{Introduction}
Recently, the LHCb collaboration performed the first full amplitude analysis of the decays $B^+\to J/\psi \phi K^+$ with $J/\psi\to\mu^+\mu^-$, $\phi\to K^+K^-$   with a data sample of 3 fb$^{-1}$ of $pp$ collision data collected at $\sqrt{s}=7$ and $8$ TeV with the LHCb detector,   confirmed the two old particles $X(4140)$ and $X(4274)$ in the $J/\psi \phi$   mass spectrum   with statistical significance $8.4\sigma$ and $6.0\sigma$, respectively,   determined the    quantum numbers  to be $J^{PC} =1^{++}$ with statistical significance $5.7\sigma$ and $5.8\sigma$, respectively \cite{LHCb-4500-1606}. Moreover, the LHCb collaboration observed  two new particles $X(4500)$ and $X(4700)$ in the $J/\psi \phi$ mass spectrum  with statistical significance $6.1\sigma$ and $5.6\sigma$, respectively,  determined the    quantum numbers  to be $J^{PC} =0^{++}$ with statistical significance $4.0\sigma$ and $4.5\sigma$, respectively \cite{LHCb-4500-1606}. The measured  masses and widths are
\begin{flalign}
 & X(4140) : M = 4146.5 \pm 4.5 ^{+4.6}_{-2.8} \mbox{ MeV}
\, , \, \Gamma = 83 \pm 21 ^{+21}_{-14} \mbox{ MeV} \, , \nonumber\\
 & X(4274) : M = 4273.3 \pm 8.3 ^{+17.2}_{-3.6} \mbox{ MeV}
\, , \, \Gamma = 56 \pm 11 ^{+8}_{-11} \mbox{ MeV} \, ,\nonumber \\
 & X(4500) : M = 4506 \pm 11 ^{+12}_{-15} \mbox{ MeV} \, ,
\, \Gamma = 92 \pm 21 ^{+21}_{-20} \mbox{ MeV} \, , \nonumber\\
 & X(4700) : M = 4704 \pm 10 ^{+14}_{-24} \mbox{ MeV} \, ,
\, \Gamma = 120 \pm 31 ^{+42}_{-33} \mbox{ MeV} \, .
\end{flalign}
There have been several possible assignments for the two new particles $X(4500)$ and $X(4700)$.

In Ref.\cite{HXChen1606}, Chen et al study the newly observed $X(4500)$  and $X(4700)$  based on the diquark-antidiquark configuration within the framework of QCD sum rules, and  interpret them as the D-wave $cs\bar{c}\bar{s}$   tetraquark states with $J^P =0^+$.

In Ref.\cite{Liu1607}, Liu studies the possible rescattering effects contribute to the process $B^+\to J/\psi\phi K^+$, and observes
  that the $D_{s}^{*+}D_{s}^-$   rescattering via the open-charmed meson loops and the $\psi^\prime \phi$
  rescattering via the $\psi^\prime K_1$  loops may simulate the structures of the $X(4140)$  and $X(4700)$, respectively, and it is hard to attribute the $X(4274)$  and $X(4500)$  to the P-wave threshold rescattering effects.

  In Ref.\cite{Maiani1607},  Maiani,  Polosa and Riquer assign the $X(4500)$ and $X(4700)$ to be the 2S tetraquark  states based on the constituent diquark model.
  Also in Ref.\cite{Zhu1607},  Zhu assigns the $X(4500)$ and $X(4700)$ to be the 2S tetraquark  states based on the constituent diquark model.

In Ref.\cite{Lebed-3915},  Lebed and Polosa  propose that the  $X(3915)$ is the ground state
$cs\bar c \bar s$ state based on  lacking of the observed $D\bar D$ and $D^*\bar{D}^*$
decays, and attribute  the single known decay mode $J/\psi \omega$ to the $\omega-\phi$ mixing effect.

The diquarks $\varepsilon^{ijk} q^{T}_j C\Gamma q^{\prime}_k$ in color
 antitriplet have  five  structures  in Dirac spinor space, where $C\Gamma=C\gamma_5$, $C$, $C\gamma_\mu \gamma_5$,  $C\gamma_\mu $ and $C\sigma_{\mu\nu}$ for the scalar, pseudoscalar, vector, axialvector  and  tensor diquarks, respectively.
In Ref.\cite{Wang1606}, we  study the   masses and pole residues of the $X(3915)$, $X(4500)$ and $X(4700)$ in the scenario of tetraquark states with the QCD sum rules  by calculating the contributions of the vacuum condensates up to dimension 10. The theoretical calculations  support  assigning   the  $X(3915)$ and $X(4500)$ to be the 1S and 2S $C\gamma_\mu\otimes\gamma^\mu C$ type scalar $cs\bar{c}\bar{s}$ tetraquark states, respectively, and assigning  the $X(4700)$ to be the 1S $C\gamma_\mu\gamma_5\otimes\gamma_5\gamma^\mu C$ type scalar $cs\bar{c}\bar{s}$ tetraquark state. In subsequent work, we take the $X(4140)$ as the diquark-antidiquark type   $cs\bar{c}\bar{s}$ tetraquark state with $J^{PC}=1^{++}$,  and  study the mass and pole residue with the QCD sum rules in details by constructing two types interpolating currents. The numerical results $M_{X_{L,+}}=3.95\pm0.09\,\rm{GeV}$ and $M_{X_{H,+}}=5.00\pm0.10\,\rm{GeV}$
disfavor assigning the $X(4140)$ to be the  $C\gamma_5\otimes \gamma_\mu C+C\gamma_\mu\otimes \gamma_5 C$ type or $C\otimes\gamma_5 \gamma_\mu C+C\gamma_\mu\gamma_5\otimes   C$ type  tetraquark state with $J^{PC}=1^{++}$ \cite{Wang1607}.

 The
attractive interactions of one-gluon exchange  favor  formation of
the diquarks in  color antitriplet $\overline{3}_{ c}$, flavor
antitriplet $\overline{3}_{ f}$ and spin singlet $1_s$ or flavor
sextet  $6_{ f}$ and spin triplet $3_s$ \cite{One-gluon}.  The calculations based on the QCD sum rules also indicate that  the favored configurations are the $C\gamma_5$ and $C\gamma_\mu$ diquark states \cite{WangDiquark,WangLDiquark}, and the heavy-light $C\gamma_5$ and $C\gamma_\mu$ diquark states have almost  degenerate masses  \cite{WangDiquark}.
In Ref.\cite{Wang-Scalar}, we construct  the $C\gamma_5\otimes \gamma_5C$, $C\gamma_\mu\otimes \gamma^\mu C$, $C\gamma_\mu\gamma_5\otimes \gamma_5\gamma^\mu C$ type interpolating currents to study the scalar tetraquark states  with the QCD sum rules in a systematic way, and observe that  the $C\gamma_5\otimes \gamma_5C$ type and $C\gamma_\mu\otimes \gamma^\mu C$ type scalar $cs\bar{c}\bar{s}$ tetraquark states have almost degenerate masses, about $4.44\,\rm{GeV}$.

The value $4.44\,\rm{GeV}$ is not robust as the masses are extracted from the QCD spectral densities at the energy scale $\mu=1\,\rm{GeV}$.
In Refs.\cite{Wang-4660-2014,WangTetraquarkCTP,Wang-Huang-NPA-2014},   we    explore the energy scale dependence of the masses $M_{X/Y/Z}$ of the hidden charm (bottom) tetraquark states   in details for the first time, and suggest a  formula,
\begin{eqnarray}
\mu&=&\sqrt{M^2_{X/Y/Z}-(2{\mathbb{M}}_Q)^2} \, ,
 \end{eqnarray}
 with the effective heavy quark mass ${\mathbb{M}}_Q$ to determine the energy scales of the  QCD spectral densities in the QCD sum rules, which works well.

Now we take a short digression to discuss the energy scale dependence of the QCD sum rules for the hidden charm or hidden bottom tetraquark states.  The correlation functions $\Pi(p)$ can be written as
\begin{eqnarray}
\Pi(p)&=&\int_{4m^2_Q(\mu)}^{s_0} ds \frac{\rho_{QCD}(s,\mu)}{s-p^2}+\int_{s_0}^\infty ds \frac{\rho_{QCD}(s,\mu)}{s-p^2} \, ,
\end{eqnarray}
through dispersion relation at the QCD side, where the $s_0$ are continuum threshold parameters. The $\Pi(p)$ are energy scale independent,
\begin{eqnarray}
\frac{d}{d\mu}\Pi(p)&=&0\, ,
\end{eqnarray}
which does not mean
\begin{eqnarray}
\frac{d}{d\mu}\int_{4m^2_Q(\mu)}^{s_0} ds \frac{\rho_{QCD}(s,\mu)}{s-p^2}\rightarrow 0 \, ,
\end{eqnarray}
 due to the following two reasons inherited from the QCD sum rules:\\
$\bullet$ Perturbative corrections are neglected, the higher dimensional vacuum condensates are factorized into lower dimensional ones therefore  the energy scale dependence of the higher dimensional vacuum condensates is modified;\\
$\bullet$ Truncations $s_0$ set in, the correlation between the threshold $4m^2_Q(\mu)$ and continuum threshold $s_0$ is unknown. \\

In the QCD sum rules for the hidden charm or hidden bottom  tetraquark states, the integrals
 \begin{eqnarray}
 \int_{4m^2_Q(\mu)}^{s_0} ds \,\rho_{QCD}(s,\mu)\,\exp\left(-\frac{s}{T^2} \right)\, ,
 \end{eqnarray}
are sensitive to the heavy quark masses $m_Q(\mu)$ or the energy scales $\mu$, where   the $T^2$ denotes
the Borel parameters. Variations of the heavy quark masses or the energy scales $\mu$ lead to changes of integral ranges $4m_Q^2(\mu)-s_0$ of the variable
$ds$ besides the QCD spectral densities $\rho_{QCD}(s,\mu)$, therefore changes of the Borel windows and predicted masses and pole residues.
We cannot obtain energy scale independent QCD sum rules, but we have an energy scale formula to determine the energy scales consistently.

According to the formula, the energy scale $\mu=1\,\rm{GeV}$  is too low to result in  robust predictions  \cite{Wang-Scalar}.
In Ref.\cite{Wang1606}, we use the energy scale formula in Eq.(2) to determine the energy scales of the QCD spectral densities in the QCD sum rules, and
observe  that  the  $X(3915)$ and $X(4500)$ can be assigned to be the 1S and 2S $C\gamma_\mu\otimes\gamma^\mu C$  type scalar $cs\bar{c}\bar{s}$ tetraquark states, respectively, and the $X(4700)$ can be assigned to be the 1S  $C\gamma_\mu\gamma_5\otimes\gamma_5\gamma^\mu C$  type scalar $cs\bar{c}\bar{s}$ tetraquark state.
In Ref.\cite{Wang1606}, we  obtain the values $M_{X(3915)}=3.91^{+0.21}_{-0.17}\,\rm{GeV}$,  $M_{X(4500)}=4.50^{+0.08}_{-0.09}\,\rm{GeV}$  and $M_{X(4700)}=4.70^{+0.08}_{-0.09}\,\rm{GeV}$. The value $M_{X(3915)}=3.91^{+0.21}_{-0.17}\,\rm{GeV}$ is much smaller than the value $4.44\,\rm{GeV}$ obtained in Ref.\cite{Wang-Scalar}.
If the $C\gamma_5\otimes \gamma_5C$ type and $C\gamma_\mu\otimes \gamma^\mu C$ type tetraquark states have degenerate masses, then the $X(3915)$ and $X(4500)$ can have another diquark-antidiquark structure, $C\gamma_5\otimes \gamma_5C$.

 In this article, we study the   $C\gamma_5\otimes \gamma_5C$ type and $C\otimes C$ type scalar $cs\bar{c}\bar{s}$ tetraquark states with the QCD sum rules  by calculating the contributions of the vacuum condensates up to dimension 10 in a consistent way. In calculations, we use the energy scale formula to determine the optimal energy scales of the QCD spectral densities to extract to tetraquark masses to identify the $X(3915)$, $X(4500)$ and $X(4700)$.

The article is arranged as follows:  we derive the QCD sum rules for the masses and pole residues of  the 1S  $C\gamma_5\otimes \gamma_5C$ type and $C\otimes C$ type $cs\bar{c}\bar{s}$ tetraquark states  in section 2; in section 3, we derive the QCD sum rules for the masses and pole residues of  the 1S  and 2S $C\gamma_5\otimes \gamma_5C$ type  $cs\bar{c}\bar{s}$ tetraquark states; section 4 is reserved for our conclusion.

\section{QCD sum rules for  the scalar tetraquark states $X_L$ and $X_H$ }
In the following, we write down  the two-point correlation functions $\Pi_{L/H}(p)$ in the QCD sum rules,
\begin{eqnarray}
\Pi_{L/H}(p)&=&i\int d^4x e^{ip \cdot x} \langle0|T\left\{J_{L/H}(x)J_{L/H}^{\dagger}(0)\right\}|0\rangle \, ,
\end{eqnarray}
where
\begin{eqnarray}
   J_L(x)&=&\epsilon^{ijk}\epsilon^{imn}s^j(x)C\gamma_5 c^k(x) \bar{s}^m(x)\gamma_5 C \bar{c}^n(x) \, , \nonumber\\
   J_H(x)&=&\epsilon^{ijk}\epsilon^{imn}s^j(x)C c^k(x) \bar{s}^m(x) C \bar{c}^n(x)\, ,
\end{eqnarray}
where the $i$, $j$, $k$, $m$, $n$ are color indexes, the $C$ is the charge conjugation matrix. The scalar diquark states are more stable than the pseudoscalar diquark states, so we expect that the $C\gamma_5\otimes \gamma_5C$ type tetraquark states have much smaller masses than the corresponding $C\otimes C$ type tetraquark states, and add the subscripts $L$ and $H$ to denote the light and heavy tetraquark states, respectively.

At the phenomenological  side, we can insert  a complete set of intermediate hadronic states with
the same quantum numbers as the current operators  $J_{L/H}(x)$ into the
correlation functions  $\Pi_{L/H}(p)$ to obtain the hadronic representation
\cite{SVZ79,Reinders85}. After isolating the ground state
contributions of the scalar $cs\bar{c}\bar{s}$ tetraquark states $X_{L/H}$, we get the results,
\begin{eqnarray}
\Pi_{L}(p)&=&\frac{\lambda_{ X_L}^2}{M_{X_L}^2-p^2} +\cdots \, \, , \nonumber\\
\Pi_{H}(p)&=&\frac{\lambda_{ X_H}^2}{M_{X_H}^2-p^2} +\cdots \, \, ,
\end{eqnarray}
where the   pole residues  $\lambda_{X_{L/H}}$ are defined by $\langle 0|J_{L/H}(0)|X_{L/H}(p)\rangle = \lambda_{X_{L/H}}$.

 In the following,  we briefly outline  the operator product expansion for the correlation functions $\Pi_{L/H}(p)$ in perturbative QCD.  We contract the $s$  and $c$ quark fields in the correlation functions $\Pi_{L/H}(p)$ with Wick theorem, and obtain the results:
 \begin{eqnarray}
 \Pi_{L}(p)&=&i\epsilon^{ijk}\epsilon^{imn}\epsilon^{i^{\prime}j^{\prime}k^{\prime}}\epsilon^{i^{\prime}m^{\prime}n^{\prime}}\int d^4x e^{ip \cdot x}   \nonumber\\
&&{\rm Tr}\left[ \gamma_{5}C^{kk^{\prime}}(x)\gamma_{5} CS^{jj^{\prime}T}(x)C\right] {\rm Tr}\left[ \gamma_{5} C^{n^{\prime}n}(-x)\gamma_{5} C S^{m^{\prime}mT}(-x)C\right]   \, , \nonumber\\
\Pi_{H}(p)&=&i\epsilon^{ijk}\epsilon^{imn}\epsilon^{i^{\prime}j^{\prime}k^{\prime}}\epsilon^{i^{\prime}m^{\prime}n^{\prime}}\int d^4x e^{ip \cdot x}   \nonumber\\
&&{\rm Tr}\left[ C^{kk^{\prime}}(x) CS^{jj^{\prime}T}(x)C\right] {\rm Tr}\left[  C^{n^{\prime}n}(-x) C S^{m^{\prime}mT}(-x)C\right]   \, ,
\end{eqnarray}
 where the $S_{ij}(x)$ and $C_{ij}(x)$ are the full  $s$ and $c$ quark propagators, respectively \cite{Reinders85,Pascual-1984},
\begin{eqnarray}
S^{ij}(x)&=& \frac{i\delta_{ij}\!\not\!{x}}{ 2\pi^2x^4}
-\frac{\delta_{ij}m_s}{4\pi^2x^2}-\frac{\delta_{ij}\langle
\bar{s}s\rangle}{12} +\frac{i\delta_{ij}\!\not\!{x}m_s
\langle\bar{s}s\rangle}{48}-\frac{\delta_{ij}x^2\langle \bar{s}g_s\sigma Gs\rangle}{192}+\frac{i\delta_{ij}x^2\!\not\!{x} m_s\langle \bar{s}g_s\sigma
 Gs\rangle }{1152}\nonumber\\
&& -\frac{ig_s G^{a}_{\alpha\beta}t^a_{ij}(\!\not\!{x}
\sigma^{\alpha\beta}+\sigma^{\alpha\beta} \!\not\!{x})}{32\pi^2x^2} -\frac{i\delta_{ij}x^2\!\not\!{x}g_s^2\langle \bar{s} s\rangle^2}{7776} -\frac{\delta_{ij}x^4\langle \bar{s}s \rangle\langle g_s^2 GG\rangle}{27648}-\frac{1}{8}\langle\bar{s}_j\sigma^{\mu\nu}s_i \rangle \sigma_{\mu\nu} \nonumber\\
&&   -\frac{1}{4}\langle\bar{s}_j\gamma^{\mu}s_i\rangle \gamma_{\mu }+\cdots \, ,
\end{eqnarray}
\begin{eqnarray}
C_{ij}(x)&=&\frac{i}{(2\pi)^4}\int d^4k e^{-ik \cdot x} \left\{
\frac{\delta_{ij}}{\!\not\!{k}-m_c}
-\frac{g_sG^n_{\alpha\beta}t^n_{ij}}{4}\frac{\sigma^{\alpha\beta}(\!\not\!{k}+m_c)+(\!\not\!{k}+m_c)
\sigma^{\alpha\beta}}{(k^2-m_c^2)^2}\right.\nonumber\\
&&\left. +\frac{g_s D_\alpha G^n_{\beta\lambda}t^n_{ij}(f^{\lambda\beta\alpha}+f^{\lambda\alpha\beta}) }{3(k^2-m_c^2)^4}
-\frac{g_s^2 (t^at^b)_{ij} G^a_{\alpha\beta}G^b_{\mu\nu}(f^{\alpha\beta\mu\nu}+f^{\alpha\mu\beta\nu}+f^{\alpha\mu\nu\beta}) }{4(k^2-m_c^2)^5}+\cdots\right\} \, , \nonumber \\
\end{eqnarray}
\begin{eqnarray}
f^{\lambda\alpha\beta}&=&(\!\not\!{k}+m_c)\gamma^\lambda(\!\not\!{k}+m_c)\gamma^\alpha(\!\not\!{k}+m_c)\gamma^\beta(\!\not\!{k}+m_c)\, ,\nonumber\\
f^{\alpha\beta\mu\nu}&=&(\!\not\!{k}+m_c)\gamma^\alpha(\!\not\!{k}+m_c)\gamma^\beta(\!\not\!{k}+m_c)\gamma^\mu(\!\not\!{k}+m_c)\gamma^\nu(\!\not\!{k}+m_c)\, ,
\end{eqnarray}
and  $t^n=\frac{\lambda^n}{2}$, the $\lambda^n$ is the Gell-Mann matrix,  $D_\alpha=\partial_\alpha-ig_sG^n_\alpha t^n$ \cite{Reinders85}.
Then we compute  the integrals both in  coordinate space and in momentum space,  and obtain the correlation functions $\Pi_{L/H}(p)$, therefore the QCD spectral densities through dispersion relation. In this article, we take into account the vacuum condensates which are
vacuum expectations  of the operators  of the orders $\mathcal{O}( \alpha_s^{k})$ with $k\leq 1$ consistently. For the technical details, one can consult Ref.\cite{WangHuangTao}. We neglect the radiative $\mathcal{O}(\alpha_s)$ corrections for the perturbative contributions, it is a challenging or formidable work to calculate the radiative $\mathcal{O}(\alpha_s)$ corrections in the QCD sum rules for hidden charm or hidden bottom tetraquark states, though the corrections may be large in the presence of two heavy quarks, just like in the QCD sum rules for the vector and axialvector   $B_c$ mesons \cite{Wangzg-Bc}.

 Once the analytical QCD spectral densities are obtained,  we  take the
quark-hadron duality below the continuum thresholds  $s^0_{L/H}$ and perform Borel transform  with respect to
the variable $P^2=-p^2$ to obtain  the QCD sum rules:
\begin{eqnarray}
\lambda^2_{X_L}\, \exp\left(-\frac{M^2_{X_L}}{T^2}\right)= \int_{4m_c^2}^{s_L^0} ds\, \rho_L(s) \, \exp\left(-\frac{s}{T^2}\right) \, ,  \\
\lambda^2_{X_H}\, \exp\left(-\frac{M^2_{X_H}}{T^2}\right)= \int_{4m_c^2}^{s_H^0} ds\, \rho_H(s) \, \exp\left(-\frac{s}{T^2}\right) \, ,
\end{eqnarray}
where
\begin{eqnarray}
\rho_L(s)&=&\rho_{0}(s)+\rho_{3}(s) +\rho_{4}(s)+\rho_{5}(s)+\rho_{6}(s)+\rho_{7}(s) +\rho_{8}(s)+\rho_{10}(s)\, , \nonumber\\
\rho_H(s)&=&\rho_L(s)|_{m_c\to -m_c} \, ,
\end{eqnarray}

\begin{eqnarray}
\rho_{0}(s)&=&\frac{1}{512\pi^6}\int_{y_i}^{y_f}dy \int_{z_i}^{1-y}dz \, yz\, (1-y-z)^3\left(s-\overline{m}_c^2\right)^2\left(7s^2-6s\overline{m}_c^2+\overline{m}_c^4 \right)    \nonumber\\
&&+\frac{m_sm_c}{256\pi^6}\int_{y_i}^{y_f}dy \int_{z_i}^{1-y}dz \, (y+z)\, (1-y-z)^2\left(s-\overline{m}_c^2\right)^2\left(5s-2\overline{m}_c^2 \right)\, ,
\end{eqnarray}

\begin{eqnarray}
\rho_{3}(s)&=&-\frac{m_c\langle \bar{s}s\rangle}{16\pi^4}\int_{y_i}^{y_f}dy \int_{z_i}^{1-y}dz \, (y+z)(1-y-z)\left(s-\overline{m}_c^2\right)\left(2s-\overline{m}_c^2\right)  \nonumber\\
&&+\frac{m_s\langle \bar{s}s\rangle}{16\pi^4}\int_{y_i}^{y_f}dy \int_{z_i}^{1-y}dz \, yz\, (1-y-z)\left(10s^2-12s\overline{m}_c^2+3\overline{m}_c^4 \right)\nonumber\\
&&-\frac{m_sm_c^2\langle \bar{s}s\rangle}{8\pi^4}\int_{y_i}^{y_f}dy \int_{z_i}^{1-y}dz  \left( s - \overline{m}_c^2\right) \, ,
\end{eqnarray}

\begin{eqnarray}
\rho_{4}(s)&=&-\frac{m_c^2}{384\pi^4} \langle\frac{\alpha_s GG}{\pi}\rangle\int_{y_i}^{y_f}dy \int_{z_i}^{1-y}dz \left( \frac{z}{y^2}+\frac{y}{z^2}\right)(1-y-z)^3 \nonumber\\
&&\left\{ 2s-\overline{m}_c^2+\frac{s^2}{6}\delta\left(s-\overline{m}_c^2\right)\right\} \nonumber\\
&&+\frac{1}{512\pi^4} \langle\frac{\alpha_s GG}{\pi}\rangle\int_{y_i}^{y_f}dy \int_{z_i}^{1-y}dz \left( y+z\right)(1-y-z)^2 \left( 10s^2-12s\overline{m}_c^2+3\overline{m}_c^4\right)  \nonumber\\
&&-\frac{m_sm_c^3}{384\pi^4} \langle\frac{\alpha_s GG}{\pi}\rangle\int_{y_i}^{y_f}dy \int_{z_i}^{1-y}dz \left( \frac{1}{y^3}+\frac{1}{z^3}\right)\left( y+z\right)(1-y-z)^2 \left\{1+\frac{s}{2}\delta\left(s-\overline{m}_c^2\right)\right\}  \nonumber\\
&&+\frac{m_sm_c}{256\pi^4} \langle\frac{\alpha_s GG}{\pi}\rangle\int_{y_i}^{y_f}dy \int_{z_i}^{1-y}dz \left( \frac{z}{y^2}+\frac{y}{z^2}\right)(1-y-z)^2 \left(3s-2\overline{m}_c^2\right)\nonumber\\
&&+\frac{m_sm_c}{128\pi^4} \langle\frac{\alpha_s GG}{\pi}\rangle\int_{y_i}^{y_f}dy \int_{z_i}^{1-y}dz \,(1-y-z) \left(3s-2\overline{m}_c^2\right)\, ,
\end{eqnarray}

\begin{eqnarray}
\rho_{5}(s)&=&\frac{m_c\langle \bar{s}g_s\sigma Gs\rangle}{64\pi^4}\int_{y_i}^{y_f}dy \int_{z_i}^{1-y}dz  \, (y+z) \left(3s-2\overline{m}_c^2 \right) \nonumber\\
&&-\frac{m_c\langle \bar{s}g_s\sigma Gs\rangle}{64\pi^4}\int_{y_i}^{y_f}dy \int_{z_i}^{1-y}dz  \, \left( \frac{y}{z}+\frac{z}{y}\right) (1-y-z) \left(3s-2\overline{m}_c^2 \right)    \nonumber\\
&&-\frac{m_s\langle \bar{s}g_s\sigma Gs\rangle}{16\pi^4}\int_{y_i}^{y_f}dy \int_{z_i}^{1-y}dz  \,yz  \left\{2s-\overline{m}_c^2 +\frac{s^2}{6}\delta\left(s-\overline{m}_c^2 \right)\right\}    \nonumber\\
&&+\frac{m_sm_c^2\langle \bar{s}g_s\sigma Gs\rangle}{32\pi^4}\int_{y_i}^{y_f}dy     \nonumber\\
&&-\frac{m_sm_c^2\langle \bar{s}g_s\sigma Gs\rangle}{64\pi^4}\int_{y_i}^{y_f}dy \int_{z_i}^{1-y}dz  \, \left( \frac{1}{y}+\frac{1}{z}\right)  \, ,
\end{eqnarray}

\begin{eqnarray}
\rho_{6}(s)&=&\frac{m_c^2\langle\bar{s}s\rangle^2}{12\pi^2}\int_{y_i}^{y_f}dy   +\frac{g_s^2\langle\bar{s}s\rangle^2}{108\pi^4}\int_{y_i}^{y_f}dy \int_{z_i}^{1-y}dz\, yz \left\{2s-\overline{m}_c^2 +\frac{s^2}{6}\delta\left(s-\overline{m}_c^2 \right)\right\}\nonumber\\
&&-\frac{g_s^2\langle\bar{s}s\rangle^2}{512\pi^4}\int_{y_i}^{y_f}dy \int_{z_i}^{1-y}dz \, (1-y-z)\left\{ 2\left(\frac{z}{y}+\frac{y}{z} \right)\left(3s-2\overline{m}_c^2 \right)+\left(\frac{z}{y^2}+\frac{y}{z^2} \right)\right.\nonumber\\
&&\left.m_c^2\left[ 2+ s\,\delta\left(s-\overline{m}_c^2 \right)\right] \right\} \nonumber\\
&&-\frac{g_s^2\langle\bar{s}s\rangle^2}{3888\pi^4}\int_{y_i}^{y_f}dy \int_{z_i}^{1-y}dz \, (1-y-z)\left\{  3\left(\frac{z}{y}+\frac{y}{z} \right)\left(3s-2\overline{m}_c^2 \right)+\left(\frac{z}{y^2}+\frac{y}{z^2} \right)\right. \nonumber\\
&&\left.m_c^2\left[ 2+s\,\delta\left(s-\overline{m}_c^2\right)\right]+(y+z)\left[12\left(2s-\overline{m}_c^2\right) +2s^2\delta\left(s-\overline{m}_c^2\right)\right] \right\}\nonumber\\
&& -\frac{m_s m_c\langle\bar{s}s\rangle^2}{12\pi^2}\int_{y_i}^{y_f}dy     \left\{1 +\frac{s}{2}\delta\left(s-\widetilde{m}_c^2 \right)\right\}\nonumber\\
&& +\frac{m_s m_c g_s^2\langle\bar{s}s\rangle^2}{648\pi^4}\int_{y_i}^{y_f}dy     \left\{1 +\frac{s}{2}\delta\left(s-\widetilde{m}_c^2 \right)\right\}\nonumber\\
&& -\frac{m_s m_c g_s^2\langle\bar{s}s\rangle^2}{216\pi^4}\int_{y_i}^{y_f}dy  \int_{z_i}^{1-y}dz   \left\{\frac{1}{y}+\frac{1}{z}+\left( \frac{1}{y^2}+\frac{1}{z^2}\right)\frac{m_c^2}{2}\delta\left(s-\overline{m}_c^2 \right)\right\}\nonumber\\
&& -\frac{m_s m_c g_s^2\langle\bar{s}s\rangle^2}{648\pi^4}\int_{y_i}^{y_f}dy  \int_{z_i}^{1-y}dz \left( \frac{y}{z}+\frac{z}{y}\right)  \left\{1+\frac{s}{2}\delta\left(s-\overline{m}_c^2 \right)\right\}\, ,
\end{eqnarray}

\begin{eqnarray}
\rho_7(s)&=&\frac{m_c^3\langle\bar{s}s\rangle}{288\pi^2  }\langle\frac{\alpha_sGG}{\pi}\rangle\int_{y_i}^{y_f}dy \int_{z_i}^{1-y}dz \left(\frac{1}{y^3}+\frac{1}{z^3} \right)(y+z)(1-y-z)\left(1+\frac{ s}{T^2}\right) \delta\left(s-\overline{m}_c^2\right)\nonumber\\
&&-\frac{m_c\langle\bar{s}s\rangle}{96\pi^2}\langle\frac{\alpha_sGG}{\pi}\rangle\int_{y_i}^{y_f}dy \int_{z_i}^{1-y}dz \left(\frac{y}{z^2}+\frac{z}{y^2}\right)(1-y-z)  \left\{2+s\,\delta\left(s-\overline{m}_c^2\right) \right\}\nonumber\\
&&-\frac{m_c\langle\bar{s}s\rangle}{96\pi^2}\langle\frac{\alpha_sGG}{\pi}\rangle\int_{y_i}^{y_f}dy \int_{z_i}^{1-y}dz\left\{2+ s\, \delta\left(s-\overline{m}_c^2\right) \right\} \nonumber\\
&&-\frac{m_c\langle\bar{s}s\rangle}{576\pi^2}\langle\frac{\alpha_sGG}{\pi}\rangle\int_{y_i}^{y_f}dy \left\{2+ s \, \delta \left(s-\widetilde{m}_c^2\right) \right\} \nonumber\\
&&-\frac{m_s m_c^2\langle\bar{s}s\rangle}{144\pi^2  }\langle\frac{\alpha_sGG}{\pi}\rangle\int_{0}^{1}dy \int_{0}^{1-y}dz \left(\frac{z}{y^2}+\frac{y}{z^2} \right)(1-y-z)\left(1+\frac{ s}{T^2}+\frac{ 2s^2}{T^4}\right) \delta\left(s-\overline{m}_c^2\right)\nonumber\\
&&+\frac{m_s m_c^4\langle\bar{s}s\rangle}{144\pi^2T^2  }\langle\frac{\alpha_sGG}{\pi}\rangle\int_{0}^{1}dy \int_{0}^{1-y}dz \left(\frac{1}{y^3}+\frac{1}{z^3} \right) \delta\left(s-\overline{m}_c^2\right)\nonumber\\
&&-\frac{m_s m_c^2\langle\bar{s}s\rangle}{48\pi^2  }\langle\frac{\alpha_sGG}{\pi}\rangle\int_{0}^{1}dy \int_{0}^{1-y}dz \left(\frac{1}{y^2}+\frac{1}{z^2} \right) \delta\left(s-\overline{m}_c^2\right)\nonumber\\
&&+\frac{m_s \langle\bar{s}s\rangle}{64\pi^2  }\langle\frac{\alpha_sGG}{\pi}\rangle\int_{y_i}^{y_f}dy \int_{z_i}^{1-y}dz \left(y+z \right) \left\{ 1+\left( \frac{2s}{3}+\frac{s^2}{6T^2}\right)\delta\left(s-\overline{m}_c^2\right)\right\}\nonumber\\
&&-\frac{m_s m_c^2\langle\bar{s}s\rangle}{288\pi^2  }\langle\frac{\alpha_sGG}{\pi}\rangle\int_{0}^{1}dy  \left( 1+\frac{s}{T^2}\right)\delta\left(s-\widetilde{m}_c^2\right) \, ,
\end{eqnarray}

\begin{eqnarray}
\rho_8(s)&=&-\frac{m_c^2\langle\bar{s}s\rangle\langle\bar{s}g_s\sigma Gs\rangle}{24\pi^2}\int_0^1 dy \left(1+\frac{s}{T^2} \right)\delta\left(s-\widetilde{m}_c^2\right)\nonumber \\
&&+\frac{ \langle\bar{s}s\rangle\langle\bar{s}g_s\sigma Gs\rangle}{48\pi^2}\int_{0}^{1} dy \,s\,\delta\left(s-\widetilde{m}_c^2\right)
 \nonumber\\
&&+\frac{ 5m_s m_c\langle\bar{s}s\rangle\langle\bar{s}g_s\sigma Gs\rangle}{144\pi^2}\int_{0}^{1} dy\left(1+\frac{s}{T^2}+\frac{s^2}{2T^4} \right)\delta\left(s-\widetilde{m}_c^2\right)
 \nonumber\\
 &&-\frac{ m_s m_c\langle\bar{s}s\rangle\langle\bar{s}g_s\sigma Gs\rangle}{96\pi^2}\int_{0}^{1} dy \left( \frac{1-y}{y}+\frac{y}{1-y}\right)\left(1+\frac{s}{T^2} \right)\delta\left(s-\widetilde{m}_c^2\right)\, ,
\end{eqnarray}

\begin{eqnarray}
\rho_{10}(s)&=&\frac{m_c^2\langle\bar{s}g_s\sigma Gs\rangle^2}{192\pi^2T^6}\int_0^1 dy \, s^2 \, \delta \left( s-\widetilde{m}_c^2\right)
\nonumber \\
&&-\frac{m_c^4\langle\bar{s}s\rangle^2}{216T^4}\langle\frac{\alpha_sGG}{\pi}\rangle\int_0^1 dy  \left\{ \frac{1}{y^3}+\frac{1}{(1-y)^3}\right\} \delta\left( s-\widetilde{m}_c^2\right)\nonumber\\
&&+\frac{m_c^2\langle\bar{s}s\rangle^2}{72T^2}\langle\frac{\alpha_sGG}{\pi}\rangle\int_0^1 dy  \left\{ \frac{1}{y^2}+\frac{1}{(1-y)^2}\right\} \delta\left( s-\widetilde{m}_c^2\right)\nonumber\\
&&-\frac{\langle\bar{s}g_s\sigma Gs\rangle^2}{192 \pi^2T^4} \int_0^1 dy   \,s^2 \, \delta\left( s-\widetilde{m}_c^2\right)\nonumber\\
&&+\frac{ \langle\bar{s}g_s\sigma Gs\rangle^2}{128 \pi^2T^2} \int_0^1 dy   \,s\,   \delta\left( s-\widetilde{m}_c^2\right)\nonumber \\
&&+\frac{m_c^2\langle\bar{s} s\rangle^2}{216 T^6}\langle\frac{\alpha_sGG}{\pi}\rangle\int_0^1 dy \, s^2 \, \delta \left( s-\widetilde{m}_c^2\right) \nonumber\\
&&-\frac{m_sm_c \langle\bar{s}g_s\sigma Gs\rangle^2}{576\pi^2T^8}\int_0^1 dy \, s^3 \, \delta \left( s-\widetilde{m}_c^2\right)
\nonumber \\
&&-\frac{m_s m_c^3\langle\bar{s}s\rangle^2}{432T^4}\langle\frac{\alpha_sGG}{\pi}\rangle\int_0^1 dy  \left\{ \frac{1}{y^3}+\frac{1}{(1-y)^3}\right\}\left(1-\frac{s}{T^2} \right) \delta\left( s-\widetilde{m}_c^2\right)\nonumber\\
&&-\frac{m_s m_c\langle\bar{s}s\rangle^2}{144T^4}\langle\frac{\alpha_sGG}{\pi}\rangle\int_0^1 dy  \left\{ \frac{1-y}{y^2}+\frac{y}{(1-y)^2}\right\}\,s\, \delta\left( s-\widetilde{m}_c^2\right)\nonumber\\
&&+\frac{m_sm_c \langle\bar{s}g_s\sigma Gs\rangle^2}{576\pi^2T^6}\int_0^1 dy \left( \frac{1-y}{y}+\frac{y}{1-y}\right)\, s^2 \, \delta \left( s-\widetilde{m}_c^2\right)
\nonumber \\
&&-\frac{m_s m_c\langle\bar{s}s\rangle^2}{864T^8}\langle\frac{\alpha_sGG}{\pi}\rangle\int_0^1 dy  \,s^3\, \delta\left( s-\widetilde{m}_c^2\right) \, ,
\end{eqnarray}
the subscripts  $0$, $3$, $4$, $5$, $6$, $7$, $8$, $10$ denote the dimensions of the  vacuum condensates, $y_{f}=\frac{1+\sqrt{1-4m_c^2/s}}{2}$,
$y_{i}=\frac{1-\sqrt{1-4m_c^2/s}}{2}$, $z_{i}=\frac{y
m_c^2}{y s -m_c^2}$, $\overline{m}_c^2=\frac{(y+z)m_c^2}{yz}$,
$ \widetilde{m}_c^2=\frac{m_c^2}{y(1-y)}$, $\int_{y_i}^{y_f}dy \to \int_{0}^{1}dy$, $\int_{z_i}^{1-y}dz \to \int_{0}^{1-y}dz$, when the $\delta$ functions $\delta\left(s-\overline{m}_c^2\right)$ and $\delta\left(s-\widetilde{m}_c^2\right)$ appear.

 We derive    Eqs.(14-15) with respect to  $\frac{1}{T^2}$, then eliminate the
 pole residues $\lambda_{X_{L/H}}$, and  obtain the QCD sum rules for
 the masses $ M_{X_{L/H}}$ of the scalar    tetraquark states,
 \begin{eqnarray}
 M^2_{X_{L/H}}&=& -\frac{\int_{4m_c^2}^{s_0} ds\,\frac{d}{d \left(1/T^2\right)}\,\rho_{L/H}(s)\,\exp\left(-\frac{s}{T^2}\right)}{\int_{4m_c^2}^{s_0} ds \, \rho_{L/H}(s)\, \exp\left(-\frac{s}{T^2}\right)}\, .
\end{eqnarray}

The vacuum condensates are taken to be the standard values
$\langle\bar{q}q \rangle=-(0.24\pm 0.01\, \rm{GeV})^3$,  $\langle\bar{s}s \rangle=(0.8\pm0.1)\langle\bar{q}q \rangle$,
 $\langle\bar{s}g_s\sigma G s \rangle=m_0^2\langle \bar{s}s \rangle$,
$m_0^2=(0.8 \pm 0.1)\,\rm{GeV}^2$, $\langle \frac{\alpha_s
GG}{\pi}\rangle=(0.33\,\rm{GeV})^4 $    at the energy scale  $\mu=1\, \rm{GeV}$
\cite{SVZ79,Reinders85,Colangelo-00}.
The quark condensates  and mixed quark condensates  evolve with the   renormalization group equation,
 $\langle\bar{s}s \rangle(\mu)=\langle\bar{s}s \rangle(Q)\left[\frac{\alpha_{s}(Q)}{\alpha_{s}(\mu)}\right]^{\frac{4}{9}}$,
 and $\langle\bar{s}g_s \sigma Gs \rangle(\mu)=\langle\bar{s}g_s \sigma Gs \rangle(Q)\left[\frac{\alpha_{s}(Q)}{\alpha_{s}(\mu)}\right]^{\frac{2}{27}}$ \cite{Narison-book}.

We take the $\overline{MS}$ masses $m_{c}(m_c)=(1.275\pm0.025)\,\rm{GeV}$ and $m_s(\mu=2\,\rm{GeV})=(0.095\pm0.005)\,\rm{GeV}$
 from the Particle Data Group \cite{PDG}, and take into account
the energy-scale dependence of  the $\overline{MS}$ masses from the renormalization group equation,
\begin{eqnarray}
m_c(\mu)&=&m_c(m_c)\left[\frac{\alpha_{s}(\mu)}{\alpha_{s}(m_c)}\right]^{\frac{12}{25}} \, ,\nonumber\\
m_s(\mu)&=&m_s({\rm 2GeV} )\left[\frac{\alpha_{s}(\mu)}{\alpha_{s}({\rm 2GeV})}\right]^{\frac{4}{9}} \, ,\nonumber\\
\alpha_s(\mu)&=&\frac{1}{b_0t}\left[1-\frac{b_1}{b_0^2}\frac{\log t}{t} +\frac{b_1^2(\log^2{t}-\log{t}-1)+b_0b_2}{b_0^4t^2}\right]\, ,
\end{eqnarray}
  where $t=\log \frac{\mu^2}{\Lambda^2}$, $b_0=\frac{33-2n_f}{12\pi}$, $b_1=\frac{153-19n_f}{24\pi^2}$, $b_2=\frac{2857-\frac{5033}{9}n_f+\frac{325}{27}n_f^2}{128\pi^3}$,  $\Lambda=213\,\rm{MeV}$, $296\,\rm{MeV}$  and  $339\,\rm{MeV}$ for the flavors  $n_f=5$, $4$ and $3$, respectively  \cite{PDG}.

  In the four-quark system $q\bar{q}^{\prime}Q\bar{Q}$,
 the $Q$-quark serves as a static well potential and  attracts  the light quark $q$  to form a heavy diquark $\mathcal{D}$ in  color antitriplet,
while the $\bar{Q}$-quark serves  as another static well potential and attracts the light antiquark $\bar{q}^\prime$  to form a heavy antidiquark $\mathcal{\bar{D}}$ in  color triplet \cite{Wang-4660-2014,WangTetraquarkCTP,Wang-Huang-NPA-2014,WangHuangTao}.
 Then  the  $\mathcal{D}$ and $\mathcal{\bar{D}}$ attract each other  to form a compact tetraquark state \cite{Wang-4660-2014,WangTetraquarkCTP,Wang-Huang-NPA-2014,WangHuangTao},
the two heavy quarks $Q$ and $\bar{Q}$ stabilize the tetraquark state \cite{Brodsky-2014}. The tetraquark states $\mathcal{D\bar{D}}$  are characterized by the effective heavy quark masses ${\mathbb{M}}_Q$ and the virtuality $V=\sqrt{M^2_{X/Y/Z}-(2{\mathbb{M}}_Q)^2}$. It is natural to take the energy  scale $\mu=V$.
We cannot obtain energy scale independent QCD sum rules, but we have an energy scale formula to determine the energy scales consistently.
We fit  the   effective $Q$-quark mass ${\mathbb{M}}_{Q}$ to reproduce the experimental value $M_{Z_c(3900)/Z_b(10610)}$, the empirical effective $Q$-quark mass ${\mathbb{M}}_Q$ is universal in the QCD sum rules for the hidden-charm or hidden bottom tetraquark states  and embodies the net effects of the complex dynamics. We take the empirical energy scale formula and reproduce the experimental values of the  masses of the  $X(3872)$, $Z_c(3900)$,  $Z_c(4020)$, $Z_c(4025)$, $Z(4430)$, $Y(4660)$, $Z_b(10610)$  and $Z_b(10650)$ in the  scenario of  tetraquark  states  \cite{Wang-4660-2014,WangTetraquarkCTP,Wang-Huang-NPA-2014,WangHuangTao,WangEPJC1601,Wang4430}.
 In this article, we take the updated value ${\mathbb{M}}_c=1.82\,\rm{GeV}$ \cite{WangEPJC1601}.

 We search for the  Borel parameters $T^2$ and continuum threshold
parameters $s^0_{L/H}$  according to  the  four criteria:

$\bf{1_\cdot}$ Pole dominance at the phenomenological side;

$\bf{2_\cdot}$ Convergence of the operator product expansion;

$\bf{3_\cdot}$ Appearance of the Borel platforms;

$\bf{4_\cdot}$ Satisfying the energy scale formula.

Now we take a short digression to discuss how to choose the Borel parameters. At the phenomenological side of the QCD sum rules, we prefer smaller Borel parameters so as to depress the contributions of the higher excited states and continuum states and determine the upper  bound of the Borel parameters $T^2_{max}$. At the QCD side of the QCD sum rules, we prefer larger Borel parameters so as to warrant the convergence of the operator product expansion and determine the lower  bound of the Borel parameters $T^2_{min}$. In the QCD sum rules for the tetraquark states, the operator product expansion converges slowly, the $T^2_{min}$ is postponed to large value, the Borel window $T^2_{max}-T^2_{min}$ is rather small. However, the small Borel window does  exist, the Borel parameter is just a free parameter, the predicted masses and pole residues should be independent on this parameter, in other words, there appears Borel platform.

The resulting Borel parameters $T^2$ and threshold parameters $s^0_{L/H}$ are
\begin{eqnarray}
  X_L &:& T^2 = (2.7-3.1) \mbox{ GeV}^2 \, ,\, s^0_{L} = (4.4\pm0.1 \mbox{ GeV})^2 \, , \\
  X_H &:& T^2 = (5.2-5.6) \mbox{ GeV}^2 \, ,\, s^0_{H} = (6.0\pm0.1 \mbox{ GeV})^2 \, .
\end{eqnarray}
The pole contributions are
\begin{eqnarray}
  X_L &:& {\rm pole}  = (39-62) \% \, \,\, {\rm at}\, \, \, \mu = 1.40 \mbox{ GeV} \, , \\
  X_H &:& {\rm pole}  = (43-58) \% \, \,\, {\rm at}\, \, \, \mu = 4.10 \mbox{ GeV} \, ,
\end{eqnarray}
the pole dominance condition is satisfied.

The contributions come from the vacuum condensates of dimension $i$ $D_{i}$ are
\begin{eqnarray}
  X_L &:& D_{0}=(37-39)\%\, , \,\, D_{3}=(65-67)\%\, , \,\, D_{4}=1\%\, , \,\,D_{5}=-(14-17)\%\, , \,\,D_{6}=(9-11)\%\, ,\nonumber\\
   &&D_{7}=2\%\, , \,\, D_{8}=-(1-2)\%\, , \,\, D_{10}   \ll 1 \% \, \,\, {\rm at}\, \,\, \mu = 1.40 \mbox{ GeV} \, , \\
    X_H &:& D_{0}=(165-174)\%\, , \,\, D_{3}=-(69-78)\%\, , \,\, D_{4}=1\%\, , \,\,D_{5}=1\%\, , \,\,D_{6}=(2-3)\%\, ,\nonumber\\
   &&D_{7}=-1\%\, , \,\, D_{8}\ll 1\%\, , \,\, D_{10}   \ll 1 \% \, \,\, {\rm at}\, \,\, \mu = 4.10 \mbox{ GeV} \, ,
\end{eqnarray}
where $i=0,\,3,\,4,\,5,\,6,\,7,\,8,\,10$, the operator product expansion is well convergent.

 Now we take into account  uncertainties of all the input parameters, and obtain the values of the masses and pole residues of the $X_L$  and $X_H$,
\begin{eqnarray}
M_{X_L}&=&3.89\pm 0.05\,\rm{GeV} \, ,  \,\,\, {\rm Experimental\,\, value} \,\,\,\,3918.4\pm 1.9\,\rm{ MeV} \, \cite{PDG}\,   ,  \\
M_{X_H}&=&5.48\pm0.10\,\rm{GeV} \, ,  \,\,\, {\rm Experimental\,\, value} \,\,\,\,4704 \pm 10 ^{+14}_{-24} \mbox{ GeV}\, \cite{LHCb-4500-1606}\,   ,    \\
\lambda_{X_L}&=&(2.21 \pm 0.22)\times 10^{-2}\,\rm{GeV}^5 \, , \nonumber\\
\lambda_{X_H}&=&(1.98\pm0.08)\times 10^{-1}\,\rm{GeV}^5 \,   ,
\end{eqnarray}
which are also shown in Figs.1-2. In Figs.1-2,  we plot the masses and pole residues with variations
of the Borel parameters at much larger intervals   than the  Borel windows shown in Eqs.(27-28).
From Eqs.(33-34) and Figs.1-2, we can see that the energy scale formula is satisfied and there appear   platforms in the Borel windows, the uncertainties originate from the Borel parameters in the Borel windows are very small, $\delta M_{X_L}/M_{X_L},\,\delta M_{X_H}/M_{X_H}\ll 1\%$, while the uncertainties originate from the Borel parameters outside of the Borel windows are rather large compared to the central values, where the criterion $\bf{1}$ or the criterion $\bf{2}$ is not satisfied. In the Borel windows, the four criteria are all satisfied, we expect to make reliable predictions.   In calculations, we observe that the predicted masses $M_{X_L}$ and $M_{X_H}$ decrease monotonously with increase of the energy scales, the uncertainty
$\delta \mu=\pm0.1\,\rm{GeV}$ can lead to uncertainties   $\delta M_{X_L}=\pm20\,\rm{MeV}$ and $\delta M_{X_H}=\pm3\,\rm{MeV}$.

The present prediction $M_{X_L}=3.89\pm 0.05\,\rm{GeV}$ is compatible with the
 experimental value $M_{X(3915)}=3918.4\pm 1.9\,\rm{ MeV}$ \cite{PDG}, so it is reasonable to assign the $X_L$ to be the $X(3915)$. The predicted mass $M_{X_H}=5.48\pm0.10\,\rm{GeV}$ lies above the upper bound of the experimental value $M_{X(4700)} = 4704 \pm 10 ^{+14}_{-24} \mbox{ GeV}$, it is impossible to assign the $X(4700)$ to be the $C-C$ type scalar tetraquark state.
 In Ref.\cite{Wang1606}, we  obtain the values $M_{X(3915)}=3.91^{+0.21}_{-0.17}\,\rm{GeV}$,  $M_{X(4500)}=4.50^{+0.08}_{-0.09}\,\rm{GeV}$,  $M_{X(4700)}=4.70^{+0.08}_{-0.09}\,\rm{GeV}$,  which are in excellent agreement with the experimental data, and support assigning the $X(3915)$ and $X(4500)$ to be the 1S and 2S $C\gamma_\mu\otimes\gamma^\mu C$ type tetraquark states, and support assigning the  $X(4700)$ to be the 1S $C\gamma_\mu\gamma_5\otimes\gamma_5\gamma^\mu C$ type tetraquark state.

For the hidden-charm mesons, the energy gaps between the ground states and the first radial excited states are $M_{\psi^\prime}-M_{J/\psi}=589\,\rm{MeV}$,
$M_{\eta_c^\prime}-M_{\eta_c}=656\,\rm{MeV}$ \cite{PDG}, $M_{Z(4430)}-M_{Z_c(3900)}=576\,\rm{MeV}$ \cite{Wang4430}. In this article, we choose the continuum threshold parameters as $\sqrt{s^0_{L}}-M_{X_L} =  0.4\sim 0.6 \mbox{ GeV} $ and $\sqrt{s^0_{H}}-M_{X_H} =  0.4\sim 0.6 \mbox{ GeV} $, the contaminations of the radial excited states can be neglected. On the other hand, the currents  $J_{L/H}(x)$   couple potentially  to the scattering states  $ J/\psi\omega$, $J/\psi \phi$,  $D_s^{*\pm} D_s^{*\mp}$, $\cdots$, we can take into account  the contributions of the  intermediate   meson-loops.
All the renormalized self-energies  contribute  a finite imaginary part to modify the dispersion relation, we can take into account the finite width effects  by the simple replacement of the hadronic spectral densities,
\begin{eqnarray}
\delta \left(s-M^2_{L/H} \right) &\to& \frac{1}{\pi}\frac{\sqrt{s}\,\Gamma_{L/H}(s)}{\left(s-M_{L/H}^2\right)^2+s\,\Gamma_{L/H}^2(s)}\, .
\end{eqnarray}
The widths $\Gamma_{X(3915)} = 20 \pm 5 \mbox{ MeV}$ \cite{PDG} and $\Gamma_{X(4700)} = 120 \pm 31 {}^{+42}_{-33} \mbox{ MeV}$ \cite{LHCb-4500-1606} are not broad, the effects of the finite widths can be absorbed safely into the pole residues \cite{Wang1606,Wang1607,WangIJTP},  the present predictions of the masses are reasonable.

Now we can obtain the conclusion tentatively that the $X(3915)$  maybe have  both $C\gamma_\mu\otimes\gamma^\mu C$ type and $C\gamma_5\otimes\gamma_5 C$ type tetraquark components.

\begin{figure}
\centering
\includegraphics[totalheight=6cm,width=7cm]{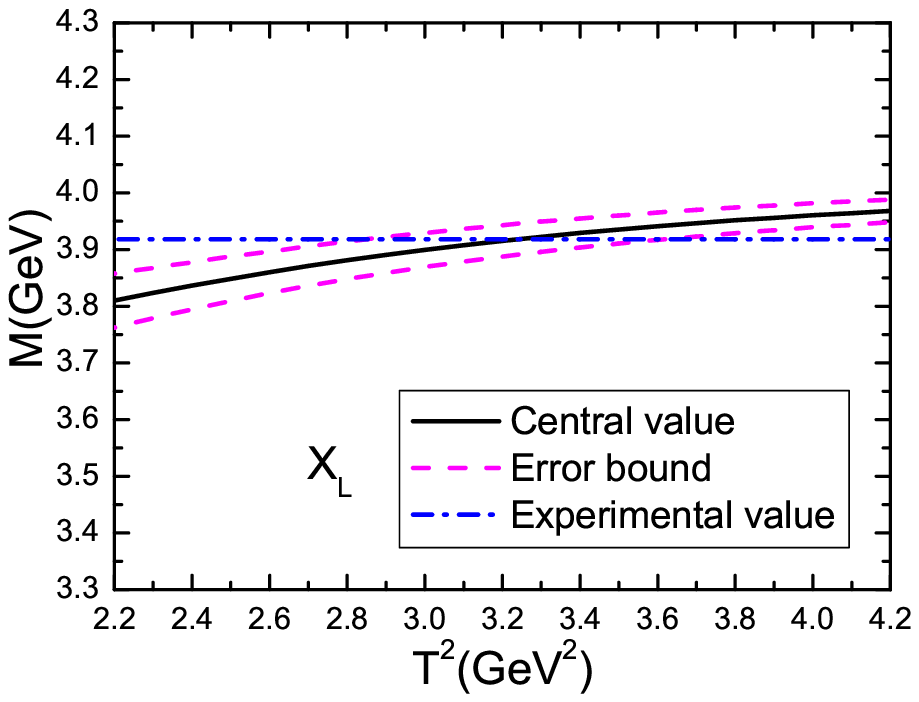}
\includegraphics[totalheight=6cm,width=7cm]{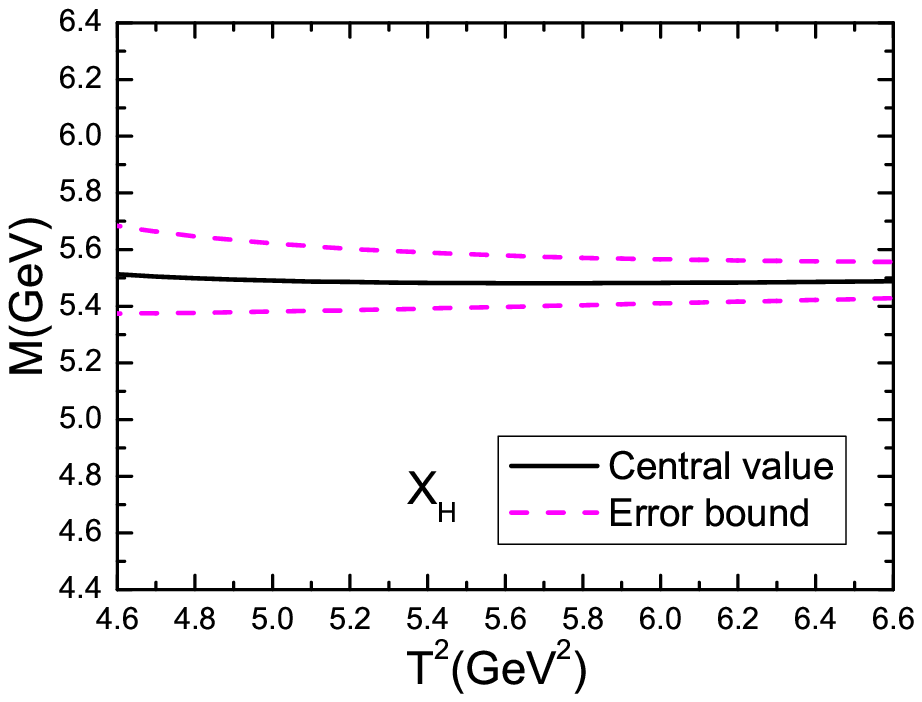}
  \caption{ The  masses $M_{X_L}$ and $M_{X_H}$  with variations of the  Borel parameters $T^2$, where the experimental value denotes the experimental value of the mass $M_{X(3915)}$. }
\end{figure}

\begin{figure}
\centering
\includegraphics[totalheight=6cm,width=7cm]{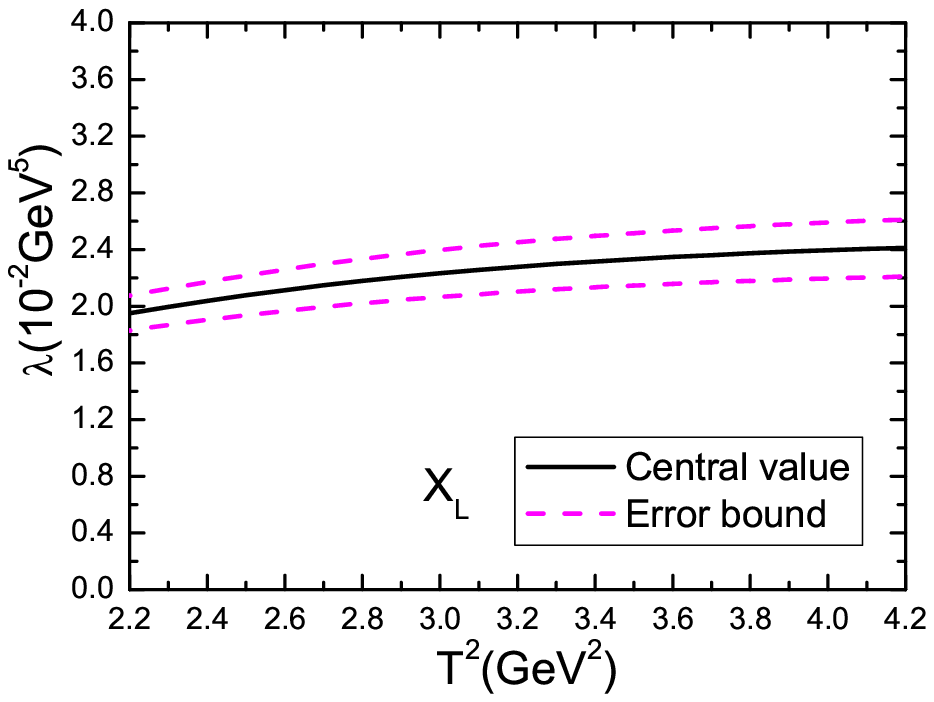}
\includegraphics[totalheight=6cm,width=7cm]{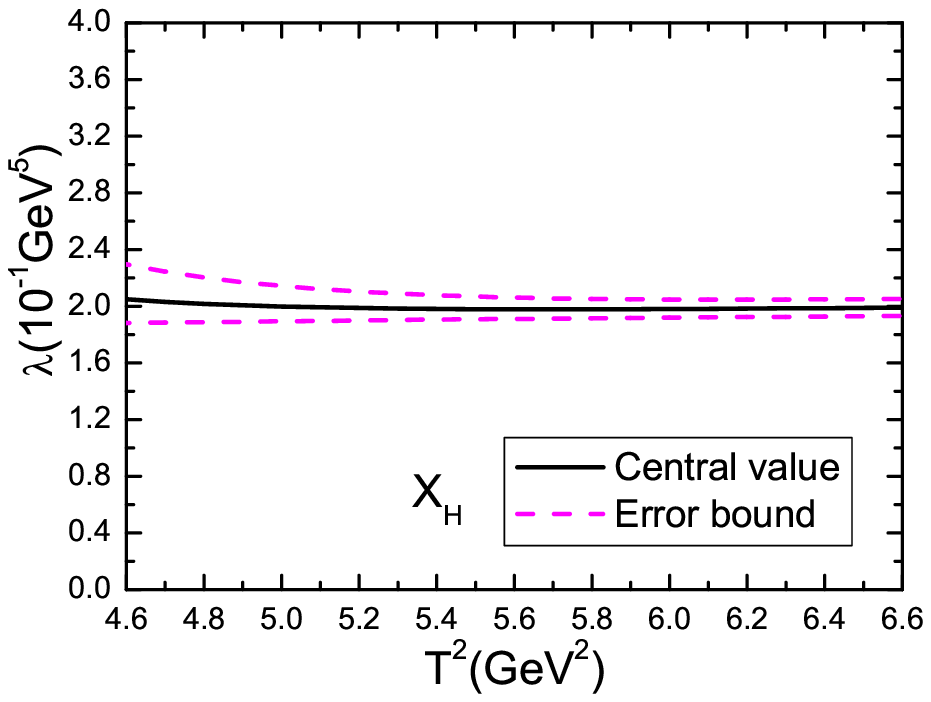}
  \caption{ The  pole residues  $\lambda_{X_L}$ and $\lambda_{X_H}$ with variations of the  Borel parameters $T^2$. }
\end{figure}

In the following, we  perform Fierz re-arrangement to the currents $J_L$ and $J_H$ both in the color and Dirac-spinor  spaces to obtain the  results,
\begin{eqnarray}
J_L&=&\frac{1}{4}\left\{\,-\bar{c} c\,\bar{s} s+\bar{c}i\gamma_5 c\,\bar{s}i\gamma_5 s-\bar{c} \gamma^\mu c\,\bar{s}\gamma_\mu s-\bar{c} \gamma^\mu\gamma_5 c\,\bar{s}\gamma_\mu\gamma_5 s+\frac{1}{2}\bar{c}\sigma_{\mu\nu} c\,\bar{s}\sigma^{\mu\nu} s\right. \nonumber\\
&&\left.+\bar{c} s\,\bar{s} c-\bar{c}i\gamma_5 s\,\bar{s}i\gamma_5 c+\bar{c} \gamma^\mu s\,\bar{s}\gamma_\mu c+\bar{c} \gamma^\mu\gamma_5 s\,\bar{s}\gamma_\mu\gamma_5 c-\frac{1}{2}\bar{c}\sigma_{\mu\nu} s\,\bar{s}\sigma^{\mu\nu} c  \,\right\} \, ,
\end{eqnarray}
\begin{eqnarray}
J_H&=&\frac{1}{4}\left\{\,\bar{c} i\gamma_5c\,\bar{s} i\gamma_5 s-\bar{c}c\,\bar{s} s+\bar{c} \gamma^\mu \gamma_5c\,\bar{s}\gamma_\mu \gamma_5s
+\bar{c} \gamma^\mu c\,\bar{s} \gamma_\mu  s+\frac{1}{2}\bar{c}\sigma_{\mu\nu} \gamma_5c\,\bar{s}\sigma^{\mu\nu}\gamma_5 s\right. \nonumber\\
&&\left.+\bar{c} s\,\bar{s} c-\bar{c}i\gamma_5 s\,\bar{s}i \gamma_5c-\bar{c} \gamma^\mu s\,\bar{s}\gamma_\mu c-\bar{c} \gamma^\mu\gamma_5 s\,\bar{s}\gamma_\mu \gamma_5c-\frac{1}{2}\bar{c}\sigma_{\mu\nu} s\,\bar{s}\sigma^{\mu\nu} c  \,\right\} \, ,
\end{eqnarray}
the components  couple  to the meson pairs, for example,  $J/\psi \phi$, $D_s\bar{D}_s$, $D_s^*\bar{D}_s^*$. The two-body strong decays
\begin{eqnarray}
X_L &\to& J/\psi \phi \to J/\psi\omega \,\,\, (\phi-\omega\,\,{\rm mixing})\, , \nonumber\\
X_H&\to& J/\psi\phi \,\, , \, \, D_{s}\bar{D}_{s}\,\, , \, \, D^*_{s}\bar{D}^*_{s}\,\, , \, \, D_{s0}\bar{D}_{s0}\,\, , \, \, D_{s1}\bar{D}_{s1}
\end{eqnarray}
are Okubo-Zweig-Iizuka  super-allowed. We can search for the $X_L$ and  $X_H$ in those decays in the future.
 The diquark-antidiquark type tetraquark state can be taken as a special superposition of a series of  meson-meson pairs, and embodies  the net effects. The decays to its components, for example, $J/\psi \phi$,  are Okubo-Zweig-Iizuka super-allowed, but the re-arrangements in the color-space are non-trivial.

\section{QCD sum rules for  the  $X(3915)$ and $X(4500)$ as the $C\gamma_5 \otimes \gamma_5C$ type tetraquark states }

Now we tentatively assign the $X(3915)$ and $X(4500)$ to be the 1S and 2S $C\gamma_5 \otimes \gamma_5C$ type tetraquark states, respectively, and  study their  masses and pole residues with the QCD sum rules.
At the phenomenological  side, we   isolate the 1S and 2S  scalar $cs\bar{c}\bar{s}$ tetraquark states in the correlation function $\Pi_{L}(p)$, and get the following  result,
\begin{eqnarray}
\Pi_{L}(p)&=&\frac{\lambda_{ X_L{(\rm 1S)}}^2}{M_{X_L{(\rm 1S)}}^2-p^2} +\frac{\lambda_{ X_L{(\rm 2S)}}^2}{M_{X_L{(\rm 2S)}}^2-p^2} +\cdots \, \, ,
\end{eqnarray}
where the   pole residues  $\lambda_{X_L{(\rm 1S,2S)}}$ are defined by $\langle 0|J_{L}(0)|X_L{(\rm 1S,2S)}(p)\rangle = \lambda_{X_L{(\rm 1S,2S)}}$.
 As the current  $J_{L}(x)$   couples  potentially  to the scattering states  $ J/\psi\omega$, $J/\psi \phi$,  $D_s^{*\pm} D_s^{*\mp}$, $\cdots$, which  contribute  a finite imaginary part to modify the dispersion relation, we can take into account the finite width effects  according to Eq.(36), and absorb the   finite width effects into the pole residues \cite{Wang1606,Wang1607,WangIJTP}. The contributions of the  scattering states  $ J/\psi\omega$, $J/\psi \phi$,  $D_s^{*\pm} D_s^{*\mp}$, $\cdots$ can be safely neglected if only the predicted masses $M_{X_L{(\rm 1S,2S)}}$ are concerned.

We take into account the contribution of the $X(4500)$ and postpone the continuum threshold $s_L^0$ to the large value $s_{X(4500)}^0$ in the QCD sum rule in Eq.(14) to obtain the QCD sum rule,
\begin{eqnarray}
\lambda^2_{X_L{(\rm 1S)}}\, \exp\left(-\frac{M^2_{X_L{(\rm 1S)}}}{T^2}\right)+\lambda^2_{X_L{(\rm 2S)}}\, \exp\left(-\frac{M^2_{X_L{(\rm 2S)}}}{T^2}\right)= \int_{4m_c^2}^{s_{X(4500)}^0} ds\, \rho_L(s) \, \exp\left(-\frac{s}{T^2}\right) \, ,
\end{eqnarray}
 then we   introduce the notations $\tau=\frac{1}{T^2}$, $D^n=\left( -\frac{d}{d\tau}\right)^n$, and use the subscripts $1$ and $2$ to denote the 1S state $X(3915)$ and the 2S state $X(4500)$ respectively for simplicity.
 The   QCD sum rule   can be rewritten as
\begin{eqnarray}
\lambda_1^2\exp\left(-\tau M_1^2 \right)+\lambda_2^2\exp\left(-\tau M_2^2 \right)&=&\Pi_{QCD}(\tau) \, ,
\end{eqnarray}
here we add the subscript $QCD$ to denote the QCD side of the correlation function $\Pi_L(\tau)$.
We derive  the QCD sum rule in Eq.(42) with respect to $\tau$ to obtain
\begin{eqnarray}
\lambda_1^2M_1^2\exp\left(-\tau M_1^2 \right)+\lambda_2^2M_2^2\exp\left(-\tau M_2^2 \right)&=&D\Pi_{QCD}(\tau) \, .
\end{eqnarray}
Then we have two equations, and obtain the QCD sum rules,
\begin{eqnarray}
\lambda_i^2\exp\left(-\tau M_i^2 \right)&=&\frac{\left(D-M_j^2\right)\Pi_{QCD}(\tau)}{M_i^2-M_j^2} \, ,
\end{eqnarray}
where $i \neq j$.
Now we derive   the QCD sum rules in Eq.(44) with respect to $\tau$ to obtain
\begin{eqnarray}
M_i^2&=&\frac{\left(D^2-M_j^2D\right)\Pi_{QCD}(\tau)}{\left(D-M_j^2\right)\Pi_{QCD}(\tau)} \, , \nonumber\\
M_i^4&=&\frac{\left(D^3-M_j^2D^2\right)\Pi_{QCD}(\tau)}{\left(D-M_j^2\right)\Pi_{QCD}(\tau)}\, .
\end{eqnarray}
 The squared masses $M_i^2$ satisfy the following equation,
\begin{eqnarray}
M_i^4-b M_i^2+c&=&0\, ,
\end{eqnarray}
where
\begin{eqnarray}
b&=&\frac{D^3\otimes D^0-D^2\otimes D}{D^2\otimes D^0-D\otimes D}\, , \nonumber\\
c&=&\frac{D^3\otimes D-D^2\otimes D^2}{D^2\otimes D^0-D\otimes D}\, , \nonumber\\
D^j \otimes D^k&=&D^j\Pi_{QCD}(\tau) \,  D^k\Pi_{QCD}(\tau)\, ,
\end{eqnarray}
$i=1,2$, $j,k=0,1,2,3$.
We solve the equation in Eq.(46) and obtain the solutions
\begin{eqnarray}
M_1^2=\frac{b-\sqrt{b^2-4c} }{2} \, , \\
M_2^2=\frac{b+\sqrt{b^2-4c} }{2} \, .
\end{eqnarray}

In Ref.\cite{Baxi-G}, M. S. Maior de Sousa and R. Rodrigues da Silva  study the masses and decay constants of the $\rho({\rm 1S,2S})$, $\psi({\rm 1S,2S})$, $\Upsilon({\rm 1S,2S})$ using   Eqs.(48-49), and observe that the ground state masses are (much) smaller than the experimental values. In Ref.\cite{Wang4430}, we apply this  approach  to study the  $Z_c(3900)$ and $Z(4430)$ as the 1S and 2S axialvector hidden-charm tetraquark states, respectively,  and use the energy scale formula
$\mu=\sqrt{M^2_{X/Y/Z}-(2{\mathbb{M}}_c)^2}$
to overcome the shortcoming \cite{Baxi-G}, and reproduce the experimental values of the masses $M_{Z_c(3900)}$ and $M_{Z(4430)}$.
In Ref.\cite{Wang1606}, we use the same  approach  to study the  $X(3915)$ and $X(4500)$ as the 1S and 2S $C\gamma_\mu\otimes\gamma^\mu C$ type scalar $cs\bar{c}\bar{s}$ tetraquark states, respectively,  and  reproduce the experimental values of the masses $M_{X(3915)}$ and $M_{X(4500)}$.
Now we tentatively  assign the   $X(3915)$ and $X(4500)$ to be the 1S and 2S $C\gamma_5 \otimes \gamma_5C$ type tetraquark states, respectively, and resort to the same approach  study the  masses and pole residues with the QCD sum rules.

Again, we search for the  Borel parameter $T^2$ and continuum threshold
parameter $s^0_{X(4500)}$  according to  the  four criteria:

$\bf{1_\cdot}$ Pole dominance at the phenomenological side;

$\bf{2_\cdot}$ Convergence of the operator product expansion;

$\bf{3_\cdot}$ Appearance of the Borel platforms;

$\bf{4_\cdot}$ Satisfying the energy scale formula.\\
The resulting Borel parameter and threshold parameter are
\begin{eqnarray}
  X(3890/4350) &:& T^2 = (1.8-2.2) \mbox{ GeV}^2 \, ,\, s^0_{X(4350)} = (4.9\pm0.1 \mbox{ GeV})^2 \, ,
  \end{eqnarray}
here we use the notations $3890$ and $4350$ in stead of $3915$ and $4500$ according to the masses extracted from the QCD sum rules.
The pole contributions are
\begin{eqnarray}
  X(3890)+X(4350) &:& {\rm pole}  = (87-98) \% \, \,\, {\rm at}\, \,\, \mu = 1.25 \mbox{ GeV} \, , \\
    X(3890)+X(4350) &:& {\rm pole}  = (93-99) \% \, \,\, {\rm at}\, \,\, \mu = 2.40 \mbox{ GeV} \, ,
\end{eqnarray}
the pole dominance condition is well satisfied.
The contributions come from the vacuum condensates of dimension 10 $D_{10}$ are
\begin{eqnarray}
  X(3890)+X(4350) &:& D_{10}  = (1-3) \% \, \,\, {\rm at}\, \,\, \mu = 1.25 \mbox{ GeV} \, , \\
    X(3890)+X(4350) &:& D_{10}  = (0-1) \% \, \,\, {\rm at}\, \,\, \mu = 2.40 \mbox{ GeV} \, ,
\end{eqnarray}
the operator product expansion is well convergent.

 Now we take into account uncertainties of all the input parameters, and obtain the masses and pole residues of the $X(3890)$ and $X(4350)$,
\begin{eqnarray}
M_{X(3890)}&=&3.85^{+0.18}_{-0.17}\,\rm{GeV} \, ,  \,\,\, {\rm Experimental\,\, value} \,\,\,\,3918.4\pm 1.9\,\rm{ MeV} \, \cite{PDG}\,   , \nonumber\\
M_{X(4350)}&=&4.83 \,\rm{GeV} \, ,  \nonumber\\
\lambda_{X(3890)}&=&1.84^{+0.76}_{-0.54}\times 10^{-2}\,\rm{GeV}^5 \, , \nonumber\\
\lambda_{X(4350)}&=&5.82\times 10^{-2}\,\rm{GeV}^5 \,   ,
\end{eqnarray}
at the energy scale $\mu=1.25\,\rm{GeV}$,
\begin{eqnarray}
M_{X(3890)}&=&3.29\,\rm{GeV} \,   , \nonumber\\
M_{X(4350)}&=&4.35^{+0.10}_{-0.11}\,\rm{GeV} \, ,  \,\,\, {\rm Experimental\,\, value} \,\,\,\,4506 \pm 11 ^{+12}_{-15}\, \rm {MeV}\, \cite{LHCb-4500-1606}\,   , \nonumber\\
\lambda_{X(3890)}&=&1.16\times 10^{-2}\,\rm{GeV}^5 \, , \nonumber\\
\lambda_{X(4350)}&=&6.01^{+0.93}_{-0.80}\times 10^{-2}\,\rm{GeV}^5 \,   ,
\end{eqnarray}
at the energy scale $\mu=2.40\,\rm{GeV}$, where we have neglected the uncertainties out of control.

\begin{figure}
\centering
\includegraphics[totalheight=9cm,width=12cm]{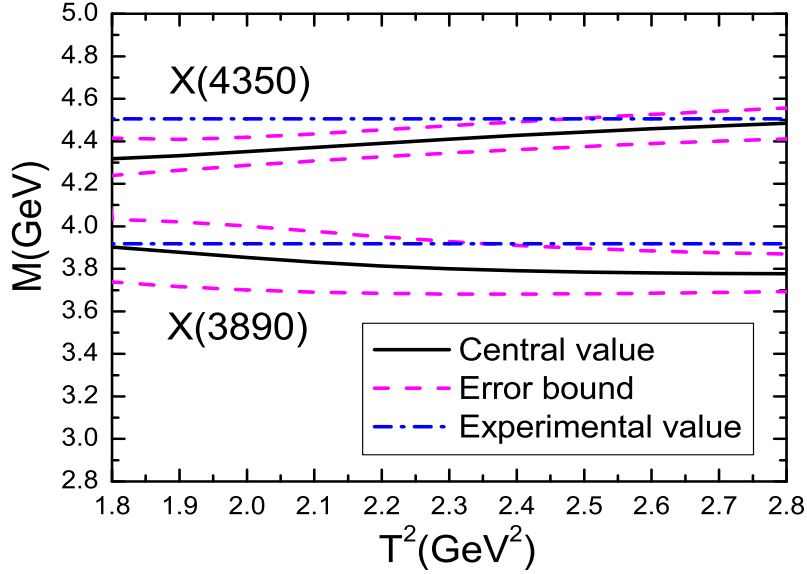}
  \caption{ The  masses $M_{X(3890)}$ and $M_{X(4350)}$ with variations of the  Borel parameters $T^2$, where the experimental value denotes the experimental values of the masses $M_{X(3915)}$ and $M_{X(4500)}$. }
\end{figure}

\begin{figure}
\centering
\includegraphics[totalheight=9cm,width=12cm]{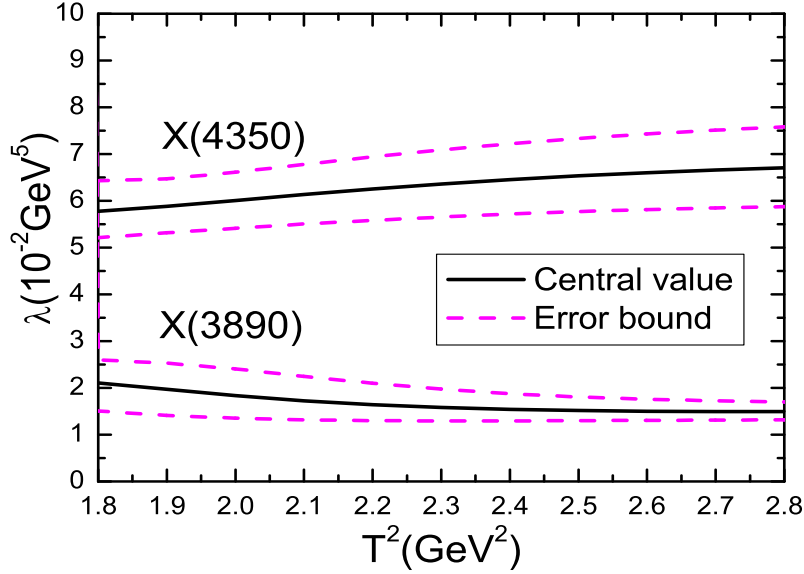}
  \caption{ The  residues  $\lambda_{X(3890)}$ and $\lambda_{X(4350)}$ with variations of the  Borel parameters $T^2$. }
\end{figure}

Then we take the central values of the masses and pole residues as the input parameters, and obtain the  pole contributions of the $X(3890)$ and $X(4350)$ respectively,
\begin{eqnarray}
{\rm pole}_{X(3890)}&=&(72-90)\% \, , \nonumber \\
{\rm pole}_{X(4350)}&=&(8-15)\% \, ,
\end{eqnarray}
at the energy scale $\mu=1.25\,\rm{GeV}$ and
\begin{eqnarray}
{\rm pole}_{X(3890)}&=&(55-76)\% \, , \nonumber \\
{\rm pole}_{X(4350)}&=&(23-38)\% \, ,
\end{eqnarray}
at the energy scale $\mu=2.40\,\rm{GeV}$.

The pole contribution of the $X(3890/4350)$  at  $\mu=1.25/2.40\,\rm{GeV}$  is larger   than that at  $\mu=2.40/1.25\,\rm{GeV}$, we prefer to extract the mass and pole residue of the $X(3890/4350)$ from the QCD spectral density  at  $\mu=1.25/2.40\,\rm{GeV}$  and neglect the ones at   $\mu=2.40/1.25\,\rm{GeV}$. The neglected values of the masses do not satisfy the energy scale formula in Eq.(2).
 In this article, we choose the  values
  \begin{eqnarray}
M_{X(3890)}&=&3.85^{+0.18}_{-0.17}\,\rm{GeV} \, ,  \,\,\, {\rm Experimental\,\, value} \,\,\,\,3918.4\pm 1.9\,\rm{ MeV} \, \cite{PDG}\,   , \nonumber\\
M_{X(4350)}&=&4.35^{+0.10}_{-0.11}\,\rm{GeV} \, ,  \,\,\, {\rm Experimental\,\, value} \,\,\,\,4506 \pm 11 ^{+12}_{-15}\, \rm {MeV}\, \cite{LHCb-4500-1606}\,   ,\\
\lambda_{X(3890)}&=&1.84^{+0.76}_{-0.54}\times 10^{-2}\,\rm{GeV}^5 \, , \nonumber\\
\lambda_{X(4350)}&=&6.01^{+0.93}_{-0.80}\times 10^{-2}\,\rm{GeV}^5 \,   ,
\end{eqnarray}
   which are also shown in Figs.3-4. In Figs.3-4,  we plot the masses and pole residues with variations
of the Borel parameter $T^2$ at much larger interval   than the  Borel window shown in Eq.(50).
There appear  platforms not very flat in the Borel window, which result in  uncertainties $\delta M_{X_L}/M_{X_L},\,\delta M_{X_H}/M_{X_H}=\pm 1\%$,  while the uncertainties originate from the Borel parameter outside of the Borel window are rather large compared to the central values.
   The predicted masses $M_{X(3890)}=3.85^{+0.18}_{-0.17}\,\rm{GeV}$ and  $M_{X(4350)}=4.35^{+0.10}_{-0.11}\,\rm{GeV}$ satisfy the energy scale formula.  From Fig.3,  we can see that the predicted mass   $M_{X(3890)}=3.85^{+0.18}_{-0.17}\,\rm{GeV}$ is compatible with the
 experimental value $M_{X(3915)}=3918.4\pm 1.9\,\rm{ MeV}$ \cite{PDG}, so it is reasonable to assign the $X(3890)$ to be the $X(3915)$. The predicted mass $M_{X(4350)}=4.35^{+0.10}_{-0.11}\,\rm{GeV}$ lies below the lower bound of the experimental value $M_{X(4500)} = 4506 \pm 11 ^{+12}_{-15}\, \rm {MeV}$ \cite{LHCb-4500-1606}, it is not favored to assign the $X(4500)$ to be the 2S $C\gamma_5\otimes \gamma_5C$ type scalar tetraquark state. Now we can draw the conclusion tentatively that the $X(3915)$ maybe have  both $C\gamma_\mu\otimes\gamma^\mu C$ type and $C\gamma_5\otimes\gamma_5 C$ type tetraquark components, while the $X(4500)$ can be assigned to the 2S $C\gamma_\mu\otimes\gamma^\mu C$ type tetraquark state \cite{Wang1606}.

In the Borel or Laplace QCD sum rules, we  introduce a new parameter $T^2$ therefore an  exponential factor $\exp\left( -\frac{s}{T^2}\right)$ to suppress the  experimentally unknown higher resonances. The Borel parameter $T^2$ has no physical significance
other than being a mathematical artifact, leads to rather   narrow stability window for the hidden charm or hidden bottom tetraquark states.
On the other hand, the continuum threshold $s_0$, which has a clear physical interpretation, is often  exponentially
suppressed $\exp\left( -\frac{s_0}{T^2}\right)$, or at best reduced in importance \cite{Dominguez-IJMPA}.  While in the finite energy QCD sum rules or  Hilbert moment QCD sum rules, the threshold $s_0$ shows  a power-like welcome feature, it is interesting to study the hidden charm or hidden bottom tetraquark states with the  finite energy QCD sum rules or  Hilbert moment QCD sum rules, this may be our next work.

\section{Conclusion}
In this article, we study the   $C\gamma_5\otimes \gamma_5C$ type and $C\otimes C$ type scalar $cs\bar{c}\bar{s}$ tetraquark states with the QCD sum rules  by calculating the contributions of the vacuum condensates up to dimension 10 in a consistent way. In calculations, we use the energy scale formula $\mu=\sqrt{M^2_{X/Y/Z}-(2{\mathbb{M}}_c)^2}$ to determine the optimal energy scales of the QCD spectral densities. The  ground state masses $M_{C\gamma_5\otimes \gamma_5C}=3.89\pm 0.05\,\rm{GeV}$  and $M_{C\otimes C}=5.48\pm0.10\,\rm{GeV}$ support assigning the $X(3915)$ to be the 1S $C\gamma_5\otimes \gamma_5C$ type tetraquark state with $J^{PC}=0^{++}$, but  do not support  assigning the $X(4700)$ to be the 1S $C\otimes C$ type $cs\bar{c}\bar{s}$ tetraquark state with $J^{PC}=0^{++}$.
 Then we tentatively assign the $X(3915)$ and $X(4500)$ to be the $\rm 1S$ and $\rm 2S$  $C\gamma_5\otimes \gamma_5C$ type  scalar $cs\bar{c}\bar{s}$ tetraquark states respectively,  and obtain the $\rm{1S}$ mass $M_{\rm 1S}=3.85^{+0.18}_{-0.17}\,\rm{GeV}$ and $\rm{2S}$ mass  $M_{\rm 2S}=4.35^{+0.10}_{-0.11}\,\rm{GeV}$ from the QCD sum rules, which support assigning the $X(3915)$ to be the $\rm{1S}$ $C\gamma_5\otimes \gamma_5C$ type tetraquark state, but  do not support  assigning the $X(4500)$ to be the ${\rm 2S}$ $C\gamma_5\otimes \gamma_5C$ type tetraquark state. The $X(3915)$ maybe have both $C\gamma_\mu\otimes\gamma^\mu C$ type and $C\gamma_5\otimes\gamma_5 C$ type tetraquark components, while the $X(4500)$ can be assigned to the 2S $C\gamma_\mu\otimes\gamma^\mu C$ type tetraquark state.

\section*{Acknowledgements}
This  work is supported by National Natural Science Foundation,
Grant Numbers 11375063,  and Natural Science Foundation of Hebei province, Grant Number A2014502017.

\end{document}